\documentclass[runningheads]{llncs}
\usepackage[T1]{fontenc}
\usepackage{graphicx}
\usepackage{url}
\begin{document}
\title{Training Users Against Human and GPT-4 Generated Social Engineering Attacks}
\titlerunning{Human and GPT-4 Social Engineering Attacks}
%
%
\author{Tailia Malloy \and
Maria José Ferreira \and 
Fei Fang \and
Cleotilde Gonzalez}
\authorrunning{Malloy et al.}
\institute{Carnegie Mellon University, Pittsburgh PA 15222, USA}
\maketitle

\begin{abstract}
    Social engineering attacks such as phishing emails remain a critical method for cybercriminals to exploit sensitive data. Although the threat of AI-generated content in such attacks is growing, current training methods predominantly rely on simplistic human-designed emails. This research introduces a novel experimental paradigm to investigate differences in the detection of human-generated versus AI-generated phishing emails, as well as two different methods by which cyberattackers could use AI as a tool to generate phishing emails. Our behavioral results reveal that emails co-created by humans and Generative-AI models pose a greater challenge to end users compared to emails created by GPT-4 or Humans working alone. We also propose a cognitive model that predicts user behavior during training, which offers the potential to be used in future user training to improve training outcomes. Our work contributes by (1) identifying critical weaknesses in current social engineering training, (2) describing biases that human participants demonstrate when viewing GPT-4 written content in emails, and (3) proposing a cognitive model-driven solution to better train users against evolving threats.
\end{abstract}

\section{Introduction}
Social engineering attacks are commonly used by cyber criminals to gain access to valuable and sensitive data. Recent Large Language Models (LLMs) such as GPT-4 have demonstrated the ability to produce convincing text that mimics human writing, and code that could be used to create fake emails and websites that appear to be legitimate \cite{achiam2023gpt}. Research in cybersecurity has identified the risks of increased proliferation of social engineering attacks through the use of LLMs \cite{schmitt2024digital}. However, the efficacy of using LLM-generated emails in training users against social engineering attacks has not been evaluated. Many training programs are based on simple human-designed emails in classroom-style instruction delivery \cite{wen2019hack}. In this work, we propose the use of GPT-4 to write convincing text that mimics real emails, and generates Javascript (JS), Hyper Text Markup Language (HTML), and Cascading Style Sheets (CSS) code to stylize emails. To our knowledge, this is the first study designed to determine the efficacy of GPT-4-generated text and code for phishing emails compared to those written by humans. We also evaluate the efficacy of emails that are written by humans and stylized with code generated by GPT-4, and vice-versa. 

Our research introduces an experimental paradigm to determine whether there is a difference in end user detection using human-written and GPT-4 generated emails. This was done in a two-by-two design that varied the original author of the email text (Human or GPT-4) as well as the style of the email (Plain-text or GPT-4 Styled). GPT-4 styled conditions indicate ones in which the emails are stylized by JS, HTML, and CSS code generated by GPT-4. A pre-experiment quiz on the indicators of phishing emails served as a measure of the base phishing knowledge of participants, and a post-experiment questionnaire had participants indicate what proportion of the content they observed was generated by AI. This allowed for an analysis of the improvement in phishing email detection, and the presence of potential biases associated with participants belief that they viewed AI-generated content. 

The results of this experiment show that emails written by humans and stylized using HTML/CSS code generated by GPT-4 are the most challenging for end users, with a significant interaction effect in which to the GPT-4 written and HTML/CSS stylized emails being the easiest for participants to categorize. Analysis of the performance of participants based on their perception of content as AI-written demonstrates a significant bias by which participants rate more emails on average as phishing if they believed a higher proportion of emails were generated by AI. This effect represents a novel \textit{AI-writing bias} that leads participants to assume that AI-written emails are phishing attempts. This bias is closely related to the well-studied phenomenon of algorithm aversion \cite{mahmud2022influences}, which has recently been demonstrated to exist in human interactions with LLMs like GPT-4 \cite{chen2024after}. Unsurprisingly, participants who had less initial knowledge of phishing emails performed worse on average under all experiment conditions compared to participants who performed better on the initial phishing quiz. 

We believe that these two groups, participants who have less initial knowledge about phishing and those who perceive all AI-written content as being more likely to be phishing, could improve their performance through a better method of selecting emails to show to participants. Such a method of improving participant training outcomes is provided in this work through a proposed Instance-Based Learning (IBL) cognitive model that uses GPT-4 embeddings of emails as attributes to predict the user's behavior. These IBL models are potentially useful in improving training outcomes by determining the best emails to show end-users during training. To evaluate this, we also run a simulation study to demonstrate how the IBL model could be used to predict the categorization of a user and, by this prediction, select an optimal email to show to that participant to optimize their training. 

\section{Background}
Generative Artificial Intelligence (GAI) has the potential to improve education and training in a variety of settings through increased accessibility and reduced costs \cite{baldassarre2023social}. However, there are significant ethical concerns due to the potential negative societal impacts of these models being misused \cite{bommasani2021opportunities}, such as through the generation of social engineering attacks \cite{al2023chatgpt}. One commonly used and widely available class of GAI is pre-trained Large Language Models (LLMs) that can be prompted to produce highly convincing textual outputs that resemble human writing \cite{sejnowski2023large}. While these methods can be trained to avoid producing potentially harmful content, these safety measures can be eschewed by repeatedly changing prompts or continuing with different prompts, in an effort to produce desired outputs \cite{white2023prompt}. The design of the prompts that are input into LLMs to produce text is call \textit{prompt engineering}, and can be used to make LLM outputs more similar to the desired output \cite{chen2023unleashing}. The repeated prompting of LLMs has been applied onto predicting how humans may speed up learning through the use of natural language instructions \cite{mcdonald2023exploring}, relating this method to approaches for training humans in different scenarios.

LLMs such as the Generative Pretrained Transformer 3 (GPT-3) \cite{brown2020language} have previously been evaluated in their social engineering ability and have shown lower performance in designing social engineering attacks compared to humans \cite{sharma2023well}. The ability of these models is constantly evolving, putting into question the ability of newer models to design social engineering attacks \cite{kumar2023certifying}. Related recent research has applied newer models like GPT-4 onto detecting phishing emails \cite{koide2024chatspamdetector}\cite{chataut2024can}. While more advanced models may be able to produce more human-like text, they also have more advanced methods to prevent misuse. This work seeks to evaluate the newer GPT-4 model \cite{achiam2023gpt} in its ability to design phishing emails, as well as to compare the effectiveness of social engineering attacks designed by humans and LLM alone and emails generated by different combinations of the output of the human and LLM model. This is an important distinction between fully LLM-generated content and content that is used in tandem with work done by cyberattackers who are leveraging AI as a tool to achieve their goals.  

Alongside this experiment, we propose a method to mitigate the potential misuse of LLMs in cybersecurity contexts by improving training against social engineering attacks. This is done by using a cognitive model to trace and predict individual learning progress and determine the best educational examples to show to participants. Optimizing educational examples can benefit participants who may have existing biases about AI-generated content, by showing them more examples of benign AI-generated content as well as potentially harmful content, so that they may learn to distinguish them. 

Overall, the contributions of this work are, first, the outline of some limitations to current social engineering training methods and, second, the identification of a potential solution to these limitations through the use of a cognitive model to improve learning outcomes. A novel bias is presented, in which participants assumed that AI-written emails are more likely to be phishing, leading to worse categorization performance. We show through simulation that selecting educational example emails using an IBL cognitive model reduces the effect of the AI-writing bias we demonstrate. These results show the usefulness of cognitive models in predicting the learning progress of end users in training scenarios, and the difficulty of correctly identifying phishing emails that are written by humans and then stylized by GPT-4.

\subsection{Large Language Models and Social Engineering Attacks}
The use of LLMs in the production of social engineering attacks demonstrates a significant concern for cybersecurity \cite{gupta2023chatgpt}. The simplicity of Generative AI tools makes them easy to apply to tasks such as writing phishing emails from scratch or stylizing existing phishing emails to look more convincing, potentially increasing their effectiveness \cite{sharma2023well}. Modern LLMs are even capable of producing code \cite{khan2022automatic}, such as JS, HTML, and CSS, \cite{lajko2022towards} that can create highly convincing emails that resemble real emails sent from many companies \cite{park2024ai}. This adds an additional layer to the potential misuse of LLMs in social engineering attacks, as hand-writing code for realistic looking emails would normally take minutes or hours, and can be done in seconds with LLMs. These two areas, writing original phishing emails and stylizing emails with HTML and CSS code, are the main focus of our experiment to investigate how users may be susceptible to social engineering attacks from humans and LLMs. 

One method of reducing the potential harm of LLMs is through the use of specific training that can make LLMs less likely to produce harmful content \cite{cao2023defending}. This is typically done using feedback from humans, either machine learning engineers or crowd-sourced participants in user studies \cite{bai2022training}. This can train models to avoid producing content that is designed to trick or scam users. However, the effectiveness of these methods in preventing the generation of dangerous content forms is not perfect and can often be worked around with more complex prompt engineering \cite{fredrikson2015model}. More advanced prompting can also train a separate model to adjust the prompt until it is accepted by the LLM and the desired content is produced \cite{zou2023universal}. In this work, we focus on using relatively simple prompt engineering to faithfully replicate what we view as a realistic scenario of a cyber attacker applying an LLM to social engineering. The prompts used to generate these emails are available in the online repository\footnote{\url{https://osf.io/wbg3r/}}, but in short they were generated by including in the prompt instructions that suggested the output would be used for educational purposes alone, which was true. 

\subsection{Social Engineering Training}
Training end users to identify social engineering attacks is an important part of cybersecurity \cite{back2021cyber}. Users without experience in security are vulnerable, making them the `weakest link' of cyber defense \cite{vishwanath2022weakest}. Phishing emails are an especially common method of social engineering due to the high volume of emails sent daily and the potential to compromise systems provided by redirecting users to unintended websites, downloading malware, or sending personal or private information, among other methods \cite{gupta2016literature}. Typically, training users to identify phishing emails focuses on specific features of these emails that can indicate that they are phishing attempts, such as the use of urgent language; making requests of confidential information; making an offer; containing a link to a dangerous website; among other features \cite{kumaraguru2009school}. In the past, this has been done using plain-text emails written by human cybersecurity experts, typically with one or more of these features included in the emails to indicate that it is a phishing attempt \cite{weaver2021training}. These training paradigms are a large industry and are commonly required by individuals, universities, companies, and other groups that are interested in improving the ability of end users to identify phishing emails \cite{jampen2020don}. Given the ever-updated nature of phishing attempts and the ease of use of LLMs in creating social engineering attacks, it is important to understand how users make decisions and learn from examples of emails written or stylized by LLMs.

The method of mitigating the potential risks associated with LLM-generated content in social engineering attacks proposed in this work involves improving the selection of training examples. The intelligence selection of training examples has previously been shown to improve student learning outcomes in domains such as geometry \cite{ferguson2006improving}, biology \cite{bouchet2013clustering}, and mathematics \cite{ritter2007cognitive}. One approach to the selection of these educational examples is called expectation maximization, which groups students based on common educational features and applies different training methods to each group \cite{woolf2010building}. Other methods attempt to improve training outcomes through the use of cognitive modeling methods to predict participant learning, and evaluate different proposed methods using simulations \cite{feng2011student}. In this work, we draw from these previous methods while designing a novel cognitive modeling approach using Instance Based Learning Theory (IBLT) \cite{gonzalez2003instance} which tracks student learning and iterates over all possible emails to determine the best email for training purposes. 

\subsection{Cognitive Modeling}
Cognitive models have previously been applied to predict human learning in anti-phishing training, demonstrating their effectiveness in accurately reflecting human learning \cite{singh2023cognitive}. Recently, Generative AI models have been integrated with cognitive models by forming \textit{representations}, of stimuli, such as textual information using LLM embeddings \cite{malloy2024applying}, \cite{malloy2024leveraging}. This approach has demonstrated human-like abilities to recognize new stimuli based on past experiences, even when they are informationally complex \cite{malloy2024efficient}. We propose the use of LLM embeddings as attributes of a cognitive model to both predict participant learning and evaluate them under different experimental conditions. These same models are also used to simulate possible improvements in phishing education that can be afforded by intelligently selecting email examples. 

\subsection{Instance Based Learning}
IBL models work by storing instances $i$ in memory $\mathcal{M}$, composed of utility outcomes $u_i$ and options $k$ composed of features $j$ in a set of features $\mathcal{F}$ of environmental decision alternatives. In the case of predicting participant learning from phishing emails, these options include labeling an email as being either dangerous (phishing) or benign (ham), the features correspond to the attributes of the email that are relevant for determining if it is a phishing email, in our model the LLM embeddings, and the outcome corresponds to the point feedback provided to participants depending on whether they are correct (1 point) or incorrect (-1 points). These options are observed in an order represented by the time step $t$, and the time step in which an instance occurred is given $\mathcal{T}(i)$. When tracing the actions of human participants, the memory is composed of the options presented to participants, the options they selected, and the utility reward that was presented to them. 

To model the retrieval of instances in memory when calculating the expected value of different option alternatives, IBL models calculate the activation of each instance in memory based on the current options available. In calculating this activation, the similarity between the instances in memory and the current instance is represented by summing over all attributes the value $S_{ij}$, which is the similarity of the attribute $j$ of the instance $i$ to the current state. This gives the activation equation as: 
 
\begin{equation}
A_i(t) = \ln \Bigg( \sum_{t' \in \mathcal{T}_i(t)} (t - t')^{-d}\Bigg) + \mu \sum_{j \in \mathcal{F}} \omega_j (S_{ij} - 1) + \sigma \xi
\label{eq:activation}
\end{equation}
The parameters of the IBL model can either be fit to individual human performance, or set to their default values. These parameters are the decay parameter $d$; the mismatch penalty $\mu$; the attribute weight of each $j$ feature $\omega_j$; and the noise parameter $\sigma$. The default values for these parameters are $(d,\mu,\omega_j,\sigma) = (0.5, 1, 1, 0.25)$. The IBL models in this work use default values to predict individual participant behaviors. The value $\xi$ is drawn from a normal distribution $\mathcal{N}(-1,1)$ and multiplied by the noise parameter $\sigma$ to add random noise to the activation. Varying these parameters impacts which instances are retrieved, and ultimately how the predicted utility of option alternatives is calculated.  

Similarities $S_{ij}$ are represented as real numbers between zero and one. In general, The modeler determines how to compute the similarity of two instances using the similarity function, which is supplied to the IBL model to be applied to values of the attributes. Different attributes can be weighted differently $w_j$ depending on their relative importance in determining the similarity of instances. The mismatch parameter $\mu$ can scale the similarity, which can be important if many instances in the decision making task are very similar or dissimilar on average. 

When predicting human learning and decision making based on textual information such as phishing emails, it is possible to use LLMs to form embeddings of these emails as attributes of the IBL model \cite{malloy2024applying}. To calculate the similarity metric $S_{ij}$ between two emails, we use the cosine similarity of their embeddings, as is done in \cite{malloy2024leveraging}. In this work, this has the benefit that the same method of forming attributes from emails can be used across experimental conditions. Thus, we can assess the effectiveness of an IBL cognitive model in predicting human learning and decision making during training. 

Once the activations of all relevant instances have been calculated, they are used to compute
the probability of retrieval $P_{i}(t)$ of the instance. This probability will determine the relevance of each instance in calculating the value of each choice option under consideration. For a given option being considered, $k$, let $\mathcal{M}_{k}$ be the set of all matching instances. Then the probability of retrieval of instance $i \in \mathcal{M}_{k}$ at time $t$ is:

\begin{equation}
P_{i}(t) = \frac{e^{A_{i}(t) / \tau}}{\sum_{i' \in \mathcal{M}_{k}}{e^{A_{i'}(t) / \tau}}}
\end{equation}

The temperature parameter $\tau$, is used in constructing this probability, and alters the selection of instances based on their activation and the relative activation of other instances. Finally, the blended value of an option $k$ is calculated at time step $t$ according to the utility outcomes $u_i$ weighted by the probability of retrieval of that instance $P_i$ and summing over all instances in memory $\mathcal{M}_k$ to give the equation:
\begin{equation}
V_k(t) = \sum_{i \in \mathcal{M}_k} P_i(t)u_i
\label{eq:blending}
\end{equation}

Where $P_i(t)$ is the probability of retrieval and $u_i$ is the utility associated with the instance $i$ in memory. There are different options for predicting the action that will be selected by humans based on the predicted values $V_k(t)$. One option is to simply maximize over the options $\mathcal{K}$ that are available to the participant $a_{t+1} = \max_{k \in \mathcal{K}} V_k(t)$. An alternative is to generate a probability distribution over actions using a temperature weighted soft-max:

\begin{equation}
p(a_{t+1},k) = \frac{e^{V_{k}(t) / \beta}}{\sum_{k' \in \mathcal{K}}{e^{V_{k'}(t) / \beta}}}
\end{equation}

This method generates a probability distribution over available actions. In the method we propose to select emails to show to participants, the next email to show is determined by first populating the model memory $\mathcal{M}$ with the experience of the participant and then iterating over all emails available $\mathcal{E}$ and finding the maximum value $V_{e'}(t)$ where $e'$ corresponds to the choice of incorrectly categorizing the email $e \in \mathcal{E}$. This gives the next email to show the participant as $e_{t+1} = \max_{e\in \mathcal{E}}V_{e'}(t)$.

\section{Experiment}
The experiment that we use is intended to train participants to identify phishing emails as dangerous, and ham emails as benign. These emails that we are interested in investigating can either be fully authored by humans, by LLMs, or a combination of the two where a human creates one of either the text body or styling, and the LLM creates the other. To test these different options of generating emails, we use a between-subjects 2x2 design varying author (Human or GPT-4) or style (plain-text or GPT-4). 

\begin{figure}[t!] 
\begin{centering}
  \includegraphics[width=\textwidth]{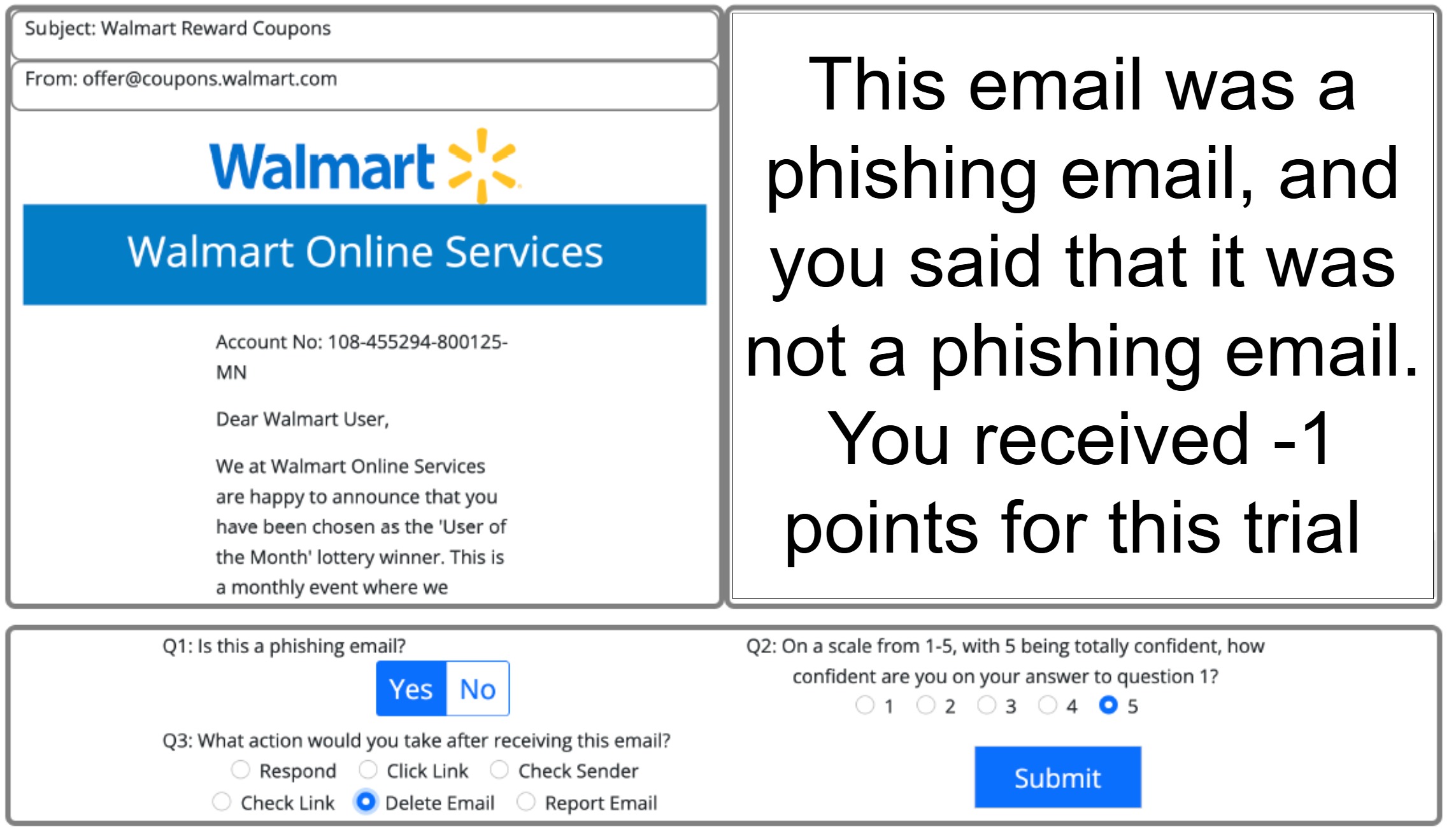} 
  \caption{An example of the email identification task shown to participants}\label{fig:Trial}
 \end{centering} 
\end{figure}

An example of the experimental interface used to evaluate the identification training of phishing emails is shown in Figure \ref{fig:Trial}. In this example, the email shown is a GPT-4 written and stylized email. Additionally, we can see that the participant in this case has incorrectly labeled the email that they were presented with as phishing, and as a result they received -1 points. While decision confidence and the action taken by the participant were collected in the experiment, only the first question determined whether participants received 1 point or -1 points. 

\begin{figure}[t!] 
\begin{centering}
  \includegraphics[width=0.9\textwidth]{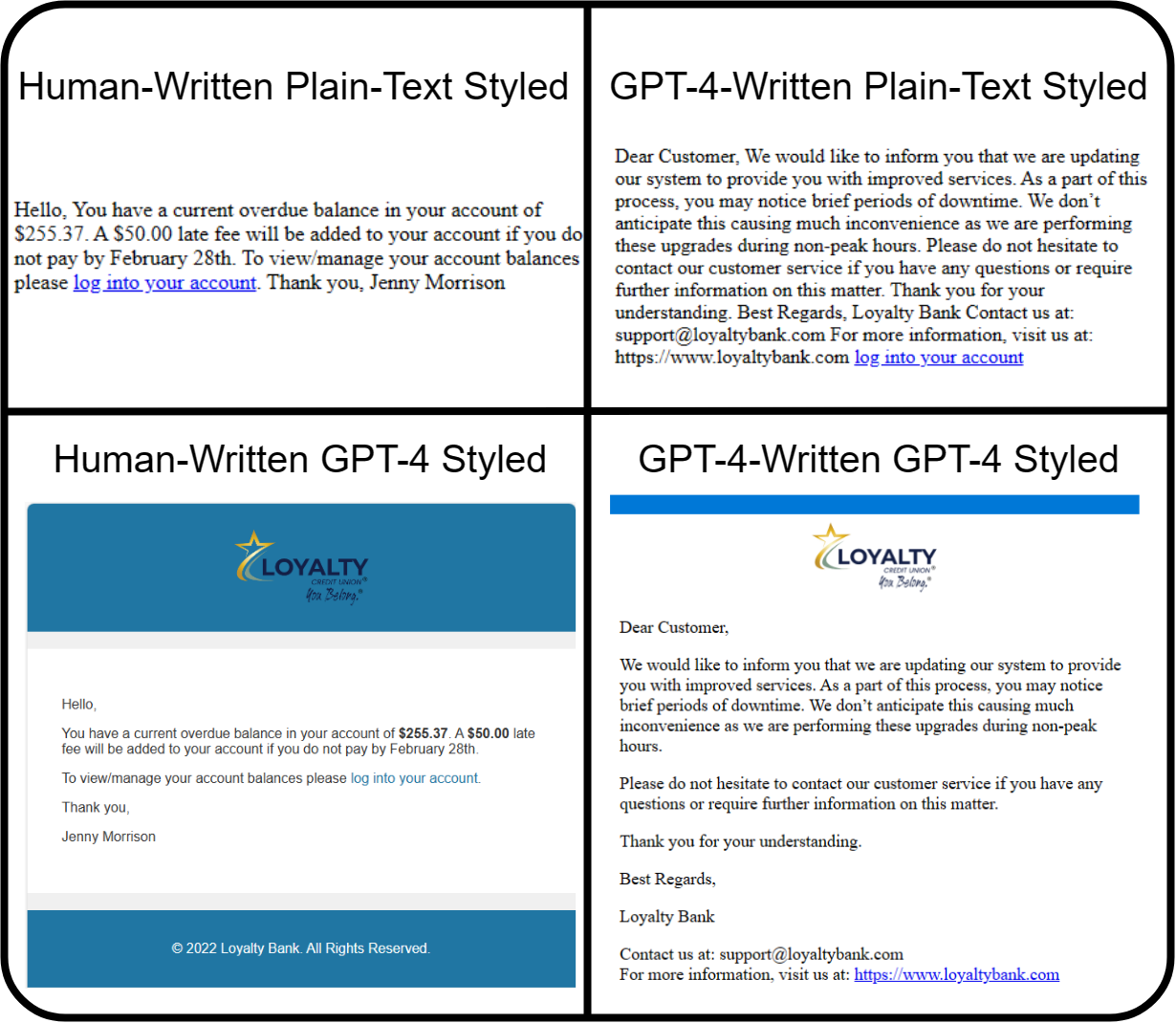} 
  \caption{Top-Left: The original plain-text email written by human experts Bottom-Left: The GPT-4 stylized version of this original email. Bottom-Right: The fully GPT-4 rewritten and stylized version of the email. Top-Right: The stripped plain-text version of the fully GPT-4 rewritten email.}\label{fig:Emails}
 \end{centering} 
\end{figure}

Another important feature of this experiment is that for each condition the same set of 360 base emails was used, all designed on alterations of an existing dataset of plain-text emails written by human cybersecurity experts that was used in a previous study \cite{singh2023cognitive}. These base emails were then either stylized by GPT-4, or rewritten entirely by prompting GPT-4 to write an email with the same attributes that the experts coded the original emails as having. The fully GPT-4 rewritten email is also stripped of HTML and CSS code and presented as the plain-text version of the GPT-4 written email. This resulted in 4 sets of 360 emails with the same general features and topics in each set. Figure \ref{fig:Emails} shows the same email that is stylized, fully rewritten, and the plain-text version of that email. 

\subsection{Methods}
This experiment compares human learning and decision making when categorizing emails as phishing (dangerous) or ham (safe) depending on the email author (Human or GPT-4) and style (plain-text or GPT-4 stylized). We are interested in determining which condition is the most difficult for humans to make accurate judgments in and whether there is a relationship between participant confidence, reaction time, and accuracy. This is an important potential relationship as it can aid in our overall goal of improving the quality of example emails shown to participants based on their performance.

This experiment included 10 pre-training trials without feedback, 40 training trials with feedback, and 10 post-training trials without feedback. During all trials, participants made judgments about emails as phishing or ham and indicated their confidence in their judgment as well as the action they would perform after reading the email. We recruited 268 participants online through the Amazon Mechanical Turk (AMT) platform. Of these participants, 44 did not complete all 60 trials and were excluded from further analysis. Of the remaining 224 participants, 18 were removed due to poor performance in the categorization task, as predefined in the study preregistration. This predefined criterion removed all participants who performed less than two standard deviations below the mean categorization improvement between pre-training and post-training trials. 

This exclusion resulted in a total of 207 participants used for the following analysis. Participants (69 Female, 137 Male, 1 Non-binary) had an average age of 40.02 with a standard deviation of 10.48 years. Of these participants, 25 had never received a phishing email, 101 had received phishing emails on a few occasions, and 79 had received phishing emails on many occasions.  Participants were compensated with a base payment of \$3 with the potential to earn up to a \$12 bonus payment depending on performance. This experiment was approved by the Carnegie Mellon University Institutional Review Board, and the study was pre-registered on OSF\footnote{\url{https://osf.io/wbg3r/}}, where all participant data and analyses are located. 

\subsection{Results}
\begin{figure}[t!] 
\begin{centering}
  \includegraphics[width=\textwidth]{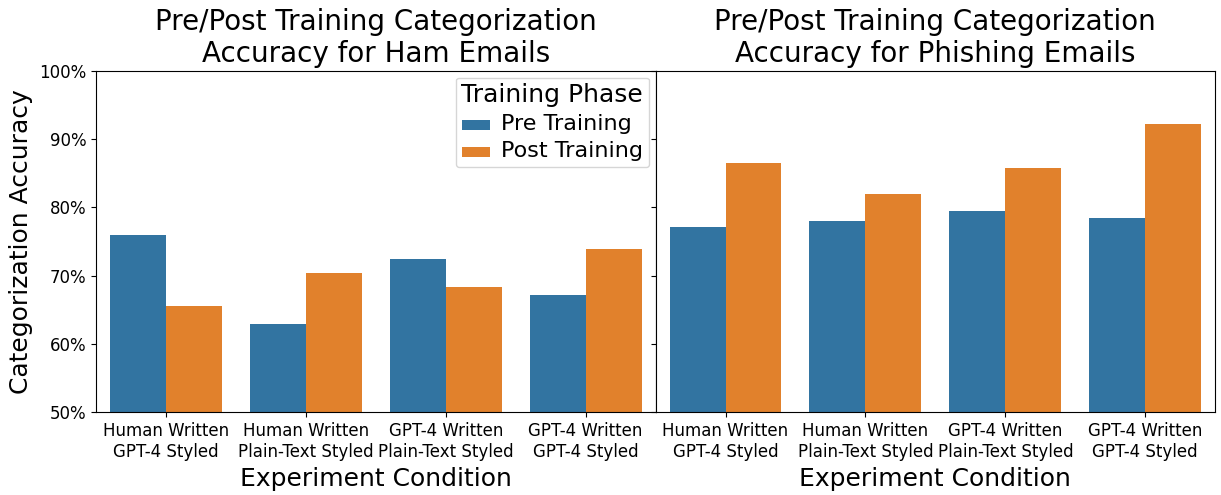} 
  \caption{Pre and post-training categorization accuracy for ham and phishing emails by experimental condition.}\label{fig:BarPerformance}
 \end{centering} 
\end{figure}

The primary comparison between conditions is done in terms of the improvement in categorization accuracy percentage between the 10 pre-training trials and the 10 post-training trials. These results are shown in Figure \ref{fig:BarPerformance}, which splits the training improvement comparison between ham and phishing emails. While phishing emails are traditionally thought of as the most relevant for training, we are also concerned of the negative outcomes that false positives can produce as more and more genuine emails we sent are being written by LLMs. For that reason we evaluate the most difficult condition for participants by taking the ham and phishing email categorization accuracy improvement to be equally relevant.

Comparing the categorization accuracy before and after training, we can see that in two conditions the categorization accuracy actually decreased, in the human written GPT-4 styled condition and the GPT-4 written plain-text styled condition for ham emails. This is an interesting result as both of the methods that combined work being done by humans and GPT-4 produced a decrease in the accuracy of ham emails. One possible reason for this is that the combination of work being done by humans and GPT-4 made the end-users uniquely apparent of the content they were observing as being AI-generated. This potential is explored further in subsequent analysis that compares the post-experiment questionnaires regarding the perception of content as being AI-generated. 

A mixed repeated measure analysis of variance of the effect of the author of the email and the style of the email on the improvement of categorization demonstrated no significant variation in author ($F=1.101,p=0.295,\eta_p^2=0.005$) but a significant variation of style ($F=12.261$, $p=0.001$, $\eta_p^2=0.057$) as well as a significant interaction between author and style ($F=14.344$, $p<0.001$, $\eta_p^2=0.066$). A post-hoc multi-comparison Tukey test showed that the improvement of the human subject in the human written and GPT-4-styled condition had a significantly lower improvement from the prior training to the post-training categorization accuracy ($p=0.033$) when compared to the GPT-4-written and GPT-4-styled condition. All other comparisons between conditions did not show a significant difference in the effect. This indicates that the smallest improvement in participant categorization accuracy was the Human written and GPT-4 styled condition ($\mu=0.015$) while the largest improvement was in the GPT-4 written and styled condition ($\mu=0.104$).

These results demonstrate the difficulty of training participants to identify emails that were written by human cybersecurity experts and stylized by GPT-4. Interestingly, the highest accuracy for the detection of phishing emails after training was observed with the written and styled by GPT-4. This is potentially due to the safety methods built into the GPT-4 model as well as the balancing of two simultaneous goals, producing the email text body and making a realistic looking email. Alternative approaches to the GPT-4 model prompting could produce more convincing phishing emails, though these complex methods may be outside of the skill set of most cybersecurity attackers. The interaction effect between the author of the email and the style may be useful to our understanding of phishing email training, since many existing platforms still use plain-text emails in training examples.

\subsection{Participant AI Identification}
To capture human participant identification of how emails were created, they were asked four questions at the end of the experiment to estimate the number of emails that they saw that were AI generated. This consisted of four questions asking what proportion of the ham and phishing emails participants believed were written by AI, and what proportion of the ham and phishing emails the participants believed were stylized by AI. Additionally, a period of 10 emails without feedback preceded and succeeded the main training experiment. This allowed us to sum together the probabilities that each participant assigned into a single value, normalized to between 0 to 100, and compare it to the difference in categorization accuracy before and after training. The next comparison we performed was to assess the average probability of a participant categorizing an email as phishing based on how likely a participant was to categorize an email as phishing based on their identification of emails as AI-generated.  

\begin{figure}[t!] 
\begin{centering}
  \includegraphics[width=\textwidth]{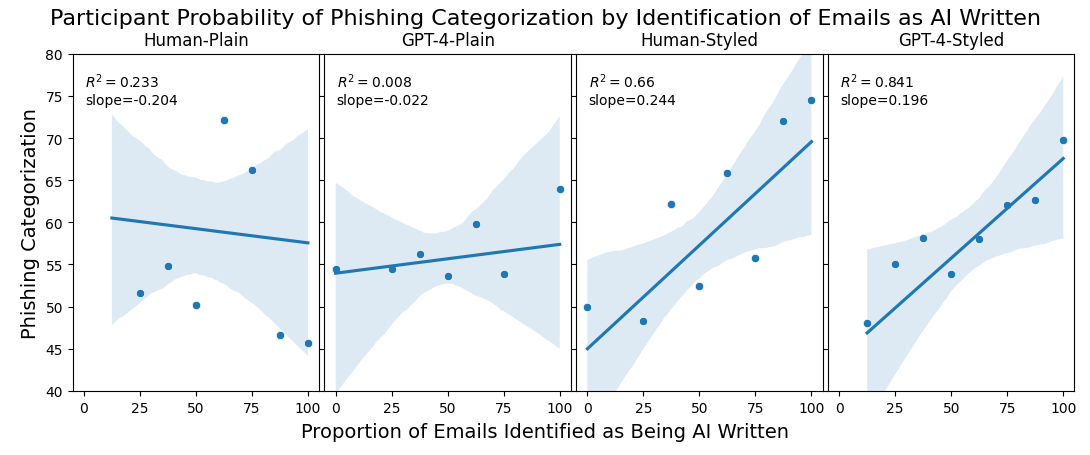} 
  \caption{Linear regression comparing the percentage of emails categorized as being phishing emails and the proportion of emails identified as being AI written. Regressions are split between each of the four experimental conditions. Shaded regions represent 95\% confidence intervals of linear regression with $R^2$ and slope labeled.}\label{fig:PerceptionCondition}
 \end{centering} 
\end{figure}

These results are shown in Figure \ref{fig:PerceptionCondition} which shows a regression of the average percent of emails classified as phishing, since half of all emails shown to the participants were phishing, a correct categorization of all emails would result in 50\% emails being classified as phishing. In general, the participants tended to categorize more than half of the emails they were shown as phishing emails. Additionally, there was an overall trend across each condition that the higher the proportion of emails identified as AI written, the higher the probability of categorizing any email as phishing. 

It may seem surprising that the increased perception of emails as written by an AI model would lead to this bias in categorizing emails as phishing. However, people generally demonstrate a poor ability to detect AI-written content \cite{kobis2021artificial}, which could interact with general aversion to algorithms \cite{burton2020systematic} which has been shown to be higher in people who have experience with algorithms making incorrect judgments \cite{dietvorst2015algorithm}. We can see from this regression that participants who identified emails as AI written in both GPT-4 style conditions were more likely to categorize emails as phishing if they had a higher identification of emails as being AI written. This represents an important bias in the identification of emails by participants that could potentially be exploited by cybersecurity attackers. This further motivates the improvement of training to detect social engineering attacks that are designed by both humans and LLMs. 

A comparison of the slopes of these regressions in Figure \ref{fig:PerceptionCondition} demonstrates that this effect of phishing categorization bias is not equal across conditions. Notably, the likelihood of categorizing emails as being phishing has both a higher slope and a higher $R^2$ for emails that were styled by GPT-4. Looking back to the four example emails shown in Figure \ref{fig:Emails}, we can see that both of the GPT-4 styled conditions include banners, logos, bold text and other styled text that may draw the attention of participants. It is likely that participants were attending to these more salient features in the GPT-4 styled conditions, which if perceived as being AI generated could bias participants into believing that emails are phishing. 

These comparisons demonstrate that there is a difference between experimental conditions in how identifying emails as being AI written impacts the likelihood of categorizing emails as being phishing. This has important implications for both understanding how participants make judgments of emails in different contexts, as well as how best to design training when incorporating LLMs into the design of example emails. It is important that participants not over attend to irrelevant features like the perception of content as being AI written, and focus on relevant features like the presence of offers or incorrect sender addresses. 

\subsection{Proposed Phishing Training supported by IBL}
Our proposed method to improve the learning outcomes of the phishing training is based on the use of an IBL model to perform model tracing during the experiment and select emails to show to participants based on that model. Specifically, this model will have the same memory of past instances, choices, and outcome observations as the individual human participant. During the pre-training and post-training trial blocks, the emails will be selected randomly from all possible emails. Then, during the training block where participants receive feedback, the model will search through all possible emails to find the email with the highest probability of being incorrectly categorized. 

\begin{figure}[!t] 
\begin{centering}
\includegraphics[width=\textwidth]{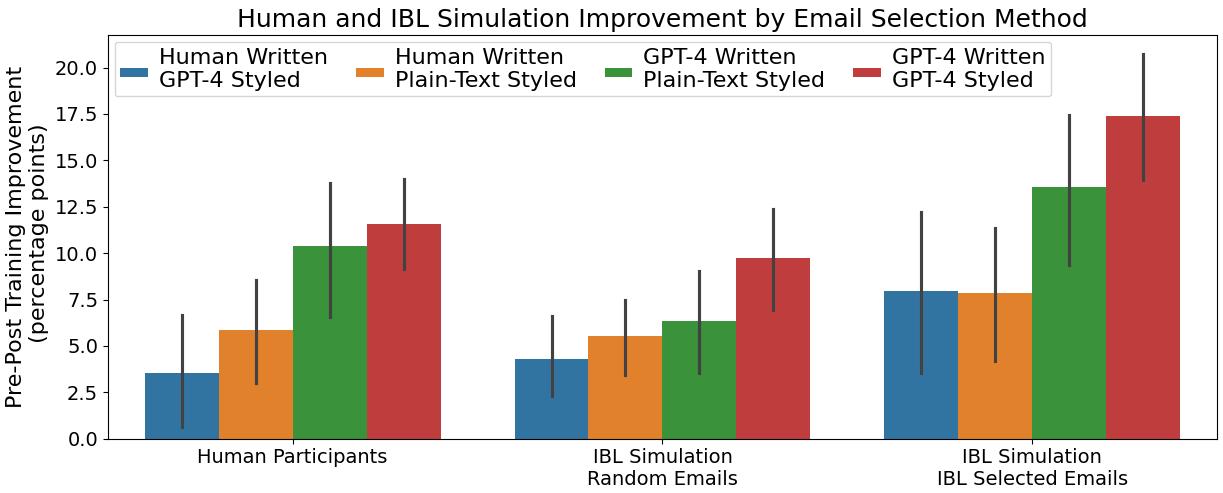} 
  \caption{All improvement measures refer to the percentage point difference between pre-training and post-training accuracy. Left: Humans participant data. Middle: Simulated IBL agents improvement under random email selection. Right: Simulated IBL agent improvement under IBL email selection method.}\label{fig:Simulations}
 \end{centering} 
\end{figure}

The theory behind this approach is that emails should be selected to show participants when there is a high probability that the participant will misclassify them. This can ensure that participants observe a diverse and challenging set of emails, based on their individual performance on past trials. This can also ensure that participants are shown similar emails when they categorize them incorrectly, until they learn the correct categorization. Since we only have data from human participants in trials in which emails are selected at random, we instead compare these two email selection training approaches using IBL models. To confirm that our IBL model simulations performed similarly as the human participants, we first compared their learning to human participants as shown in the left and middle columns of \ref{fig:Simulations}. From this, we can see that IBL model simulations produce similar improvements in performance compared to human participants, indicating that they can be helpful in our evaluation of our proposed method of selecting email training examples. 

To ensure that the IBL simulation models we are using have similar performance as human participants, their parameters were adjusted to reflect the same pre-post training improvement that was observed in humans. This can be seen in the middle column of Figure \ref{fig:Simulations}, which shows that the IBL simulated behavior has roughly the same pre-post training improvements as the human participants. The IBL simulated agents have the same training as in the experiment, with 10 pre-training trials without feedback, 40 training trials with feedback, and 10 post-training trials without feedback. 

These IBL models trained with a random selection of training emails are compared to the same IBL models trained with emails selected by a separate \textit{IBL selection}. This selection method is structured in the same way as the IBL tracing models described in previous sections. The IBL email selection method here predicts the behavior of simulated IBL learning agents, and selects the email to maximize incorrect categorization.  

The results of this training method are shown in the right column of Figure \ref{fig:Simulations}, and demonstrate a clear and significant improvement between the training results, as measured by pre-post-training improvement in terms of percentage point accuracy, between random email selection and the IBL email selection method. This suggests that selecting emails to show participants using an IBL model may improve the quality of educational outcomes. Overall, this comparison of different methods to train simulated participants provides support for our planned study that will use an IBL email selection method model to select the emails that real human participants will observe. 

\section{Discussion}
In this work, we present a method for assessing the ability of end-users to detect phishing emails written by GPT-4, humans, and through two different collaborations of GPT-4 and human work. To our knowledge this is the first experimental comparison of human participant ability to learn from these different types of phishing emails. The results of this experiment highlight issues with current methods for training end-users to identify phishing emails, namely relying on human written and plain-text emails. The is because most difficult type of email for end-users to correctly categorize in this experiment was those that were written by humans and stylized with JS, HTML and CSS code generated by GPT-4. 

Alongside this, we present a proposed solution to the issues that we highlight, to improve the quality of phishing email identification training with the aid of a cognitive model. This is done by using an Instance-Based Learning model to select the emails that are shown to participants and improve their learning outcomes. There has been a long history of research into the optimal selection of examples to show to students, and we apply similar methods onto our IBL model that uses LLM embeddings to represent emails. We motivate the applicability of this model through a simulation that estimates the potential improvement on end-user training that can be afforded by using an IBL model in the way we introduce. While these simulations are promising, additional future work is required to confirm the relevance of email selection in training outcomes. 

Several interesting and surprising results from analyses of human behavior were revealed in our experimental result. Firstly, the most significant difference between any two conditions of the experiment was in the human-written and GPT-4-styled condition and the GPT-4-written and GPT-4-styled condition. Comparing pre-training performance and improvement in the plain-text styled conditions showed little difference between different email authors. This interaction demonstrates that the GPT-4 model is unlikely to write convincing phishing emails from scratch without more advanced prompt engineering.

Another important result from the experimental analysis was the observed bias between the perception of emails as being generated by an AI model. As participants were more likely to perceive emails as being written or stylized by AI, the worse their performance in categorizing ham emails. It is possible that the presence of this bias could be incorporated into improved feedback to participants, to point out that AI generated writing does not necessarily indicate that an email is phishing. 

Improving education of AI-generated content is an important step to preventing the misuse of LLMs in the future, by improving the public awareness of the capabilities of LLMs, and how best to detect when they are potentially being used for nefarious purposes. A significant area of research in machine learning is seeking to further the capabilities of LLMs, aligning their outputs to human goals and use cases, and making misuse more difficult. However, it is unlikely that a perfect model will ever be trained, as it is possible to train separate models to learn how to best prompt LLMs to allow for unintended use cases. Thus, proper education and training are a crucial step in reducing the potential harm of LLMs in the future. 

\section*{Acknowledgments}
This research was sponsored by the Army Research Office and accomplished under Australia-US MURI Grant Number W911NF-20-S-000, and the AI Research Institutes Program funded by the National Science Foundation under AI Institute for Societal Decision Making (AI-SDM), Award No. 2229881. Compute resources and GPT model credits were provided by the Microsoft Accelerate Foundation Models Research Program grant ``Personalized Education with Foundation Models via Cognitive Modeling"

\clearpage 

\bibliography{springer}

\end{document}